\documentclass[12pt]{article}
\usepackage{fullpage,citesort,
epsfig,psfrag,graphics,amsbsy,amssymb,amsmath
}
\usepackage{caption}
\newcommand{\beq}{\begin{equation}}
\newcommand{\eeq}{\end{equation}}
\newcommand{\bea}{\begin{eqnarray}}
\newcommand{\eea}{\end{eqnarray}}
\newcommand{\bfs}{\boldsymbol}

\newcommand{\Tr}{{\rm Tr}}
\newcommand{\tr}{{\rm tr}\ }
\newcommand{\ket}[1]{|#1\rangle}

\def\math{\mathsurround=0pt }
\def\leftrightarrowfill{$\math \mathord\leftarrow \mkern-6mu
 \cleaders\hbox{$\mkern-2mu \mathord- \mkern-2mu$}\hfill
 \mkern-6mu \mathord\rightarrow$}
\def\overleftrightarrow#1{\vbox{\ialign{##\crcr
     \leftrightarrowfill\crcr\noalign{\kern-1pt\nointerlineskip}
     $\hfil\displaystyle{#1}\hfil$\crcr}}}

\newcommand{\VEV}[1]{\left\langle#1\right\rangle}

\let\l=\lambda

 \def\bd{\begin{document}} \def\ed{\end{document}}
\def\ds{\documentstyle} \let\fr=\frac \let\bl=\bigl \let\br=\bigr
\let\Br=\Bigr \let\Bl=\Bigl
\let\bm=\bibitem
\let\na=\nabla
\let\pa=\partial \let\ov=\overline
\def\ft#1#2{{\textstyle{{\scriptstyle #1}\over {\scriptstyle #2}}}}
\def\fft#1#2{{#1 \over #2}}
\def\vp{\varphi}
\def\sst#1{{\scriptscriptstyle #1}}
\def\oneone{\rlap 1\mkern4mu{\rm l}}
\def\td{\tilde}
\def\wtd{\widetilde}
\def\dalemb#1#2{{\vbox{\hrule height .#2pt
        \hbox{\vrule width.#2pt height#1pt \kern#1pt
                \vrule width.#2pt}
        \hrule height.#2pt}}}
\def\square{\mathord{\dalemb{6.8}{7}\hbox{\hskip1pt}}}
\def\wtd{\widetilde}
\def\R{\rlap{\rm I}\mkern3mu{\rm R}}
\def\im{{\rm i}}
\def\tilg{\tilde{g}}
\def\tilF{\tilde{F}}
\def\tilA{\tilde{A}}
\def\varf{\varphi}
\def\tilf{\tilde{\phi}}
\def\tilh{\tilde{h}}
\def\rme{{\rm e}}
\def\ep{\epsilon}
\def\0{{(0)}}
\def\9{{(9)}}
\def\8{{(8)}}
\def\7{{(7)}}
\def\6{{(6)}}
\def\5{{(5)}}
\def\4{{(4)}}
\def\3{{(3)}}
\def\2{{(2)}}
\def\1{{(1)}}
\newcommand{\trace}{{\rm Tr}}
\newcommand{\ub}{\overline{U}}
\newcommand{\vb}{\overline{V}}
\newcommand{\uh}{\widehat{U}}
\newcommand{\vh}{\widehat{V}}
\newcommand{\ubh}{\overline{\widehat{U}}}
\newcommand{\vbh}{\overline{\widehat{V}}}
\newcommand{\lb}{\bar{\l}}
\newcommand{\Fb}{\overline{F}}
\newcommand{\Fh}{\widehat{F}}
\newcommand{\Fbh}{\overline{\widehat{F}}}
\newcommand{\Ab}{\overline{A}}
\newcommand{\Ah}{\widehat{A}}
\newcommand{\Abh}{\overline{\widehat{A}}}
\newcommand{\Gb}{\overline{G}}
\newcommand{\Gh}{\widehat{G}}
\newcommand{\Gbh}{\overline{\widehat{G}}}
\newcommand{\Pb}{\overline{P}}
\newcommand{\Ph}{\widehat{P}}
\newcommand{\Pbh}{\overline{\widehat{P}}}
\newcommand{\Qb}{\overline{Q}}
\newcommand{\Qh}{\widehat{Q}}
\newcommand{\Qbh}{\overline{\widehat{Q}}}
\newcommand{\Bb}{\overline{B}}
\newcommand{\Bh}{\widehat{B}}
\newcommand{\Bbh}{\overline{\widehat{B}}}
\newcommand{\fhns}{\hat{F}^{\rm (NS)}}
\newcommand{\fhrr}{\hat{F}^{\rm (RR)}}
\newcommand{\ahns}{\hat{A}^{\rm (NS)}}
\newcommand{\ahrr}{\hat{A}^{\rm (RR)}}
\newcommand{\hhrr}{\hat{H}^{\rm (RR)}}
\newcommand{\hchi}{\hat{\chi}}
\newcommand{\hphi}{\hat{\phi}}
\newcommand{\htau}{\hat{\tau}}
\newcommand{\cG}{{\cal G}}
\newcommand{\cGb}{\overline{{\cal G}}}
\newcommand{\cH}{{\cal H}}
\newcommand{\cP}{{\cal P}}
\newcommand{\cPb}{\overline{{\cal P}}}
\newcommand{\cQ}{{\cal Q}}
\newcommand{\cQb}{\overline{{\cal Q}}}
\newcommand{\cM}{{\cal M}}
\newcommand{\cN}{{\cal N}}
\newcommand{\cO}{{\cal O}}
\newcommand{\cD}{{\cal D}}
\newcommand{\cL}{{\cal L}}

\newcommand{\vpp}{\mbox{$\langle{\scriptstyle++}\rangle$}}
\newcommand{\vmp}{\mbox{$\langle{\scriptstyle-+}\rangle$}}
\newcommand{\vppp}{\mbox{$\langle{\scriptstyle+++}\rangle$}}
\newcommand{\vmpp}{\mbox{$\langle{\scriptstyle-++}\rangle$}}
\newcommand{\vpmp}{\mbox{$\langle{\scriptstyle+-+}\rangle$}}

\begin{document}
\setlength{\captionmargin}{36pt}
\begin{titlepage}
\begin{flushright}
\phantom{UFIFT-HEP}
\end{flushright}

\vskip 3cm
\begin{center}
\begin{large}
{\bf String Bits at Finite Temperature and the Hagedorn Phase}
\end{large}

\vskip 2cm
{\large
Charles B. Thorn\footnote{E-mail  address: {\tt thorn@phys.ufl.edu}}
}
\vskip0.20cm
{\it Institute for Fundamental Theory\\
Department of Physics, University of Florida,
Gainesville FL 32611}


\vskip 1.0cm
\end{center}

\begin{abstract}
\noindent We study the behavior of a simple string bit model
at finite temperature. We use thermal perturbation theory to analyze
the high temperature regime. But at low temperatures we rely
on the large $N$ limit of the dynamics, for which the
exact energy spectrum is known. Since the lowest energy states at infinite
$N$ are free closed strings, the $N=\infty$ partition function
diverges above a finite temperature $\beta_H^{-1}$, the Hagedorn temperature.
We argue that in these models at finite $N$, which then have a finite number of
degrees of freedom, there can be neither
an ultimate temperature nor any kind of phase transition.
We discuss how the discontinuous behavior seen at infinite
$N$ can be removed at finite $N$. In this resolution the
fundamental string bit degrees of freedom become more active
at temperatures near and above the Hagedorn temperature.
\end{abstract}
\vfill
\end{titlepage}
\section{Introduction}
Half a century ago Hagedorn proposed an experimentally
successful statistical model of 
strongly interacting particles in which the density of states
grows exponentially with energy $d(E)\sim E^\alpha e^{\beta_H E}$,
with the thermodynamic consequence that $\beta_H^{-1}$ is the
ultimate temperature which cannot be exceeded by hadronic
matter in thermal equilibrium \cite{hagedorn}. The discovery that 
dual resonance models (a.k.a. string theory) predicted an energy level 
degeneracy with just this exponential behavior at zero coupling
\cite{fubiniveneziano} provided unanticipated
early support for string theory as a
model of strong interactions. Hagedorn's thermal interpretation of
an exponential level density has also been exploited
to apply string theory to early universe cosmology,
first to describe the role of the strong interactions in the hot early
universe \cite{huangweinberg}. Later on, after string theory was
promoted from a faulty model of strong interactions to a promising
vehicle for unifying quantum gravity with the rest of physics, 
the Hagedorn model formed the basis for string gas cosmology \cite{brandvafa},
which provides an alternative to the inflationary universe. 
Apparently string gas cosmology is still viable after all
these years \cite{brand15}.

Thinking
about the thermal properties of a system often leads to theoretical
insight into puzzling aspects of the system. For example, Atick and Witten,
motivated in part by parallels with the temperature dependence
of large $N$ QCD \cite{thornult}, interpreted the ultimate temperature
of the free string as an artifact of the zero coupling limit.
They suggested that, at finite coupling, there should be a phase transition near
the Hagedorn temperature to a new phase dominated by the
fundamental degrees of freedom underlying string theory \cite{atickwitten}.  
They argued that, much as in QCD where there are many fewer quarks and gluons  
than mesons and baryons, the true degrees of freedom of string theory
are probably much reduced compared to expectations from string
field theory.

It has been proposed that string 
should be regarded as a composite system of fundamental
entities \cite{gilest,thornsakh} called ``string bits'' \cite{thornps}.
It is then of interest to study string bit models at finite 
temperature and to explore how the string bit degrees of freedom
can be exposed at high temperature.
In string bit dynamics a string
bit is a discrete bit of lightcone parameterized \cite{goddardgrt}
string. Then
the total $P^+=(P^0+P^1)/\sqrt{2}$ of a string is discretized as ${\cal M}m$ 
where ${\cal M}$ is the bit number operator. 
The string itself is simply a long chain
of string bits whose nearest neighbor dynamics is implemented 
by introducing  $N\times N$ ``color'' matrix string bit creation operators,
imposing a $U(N)$ color symmetry. 
Then we identify  string perturbation theory as 
the 't Hooft $1/N$ expansion
\cite{thooftlargen} of string bit dynamics. In early versions of this 
dynamics \cite{thornsakh} the
creation operators were fields depending on transverse coordinates ${\bfs x}$,
as well as spinor indices in the case of superstring bits \cite{bergmantsubit}.
However, each transverse coordinate can effectively emerge from
a simple two valued internal flavor degree of freedom \cite{gilesmt}, so 
spaceless string
bit models (in zero space dimensions) can underlie string theory
\cite{sunthorn,spacebits} in any space dimension less than or equal to
the critical one. 

Since the Hagedorn phenomenon is
common to all string models, even subcritical ones, 
we choose to analyze this phenomenon in the simplest stable superstring
bit model studied in \cite{sunthorn}. 
Its large $N$ 
limit describes a noncovariant subcritical 
lightcone string with no transverse coordinates and
one Grassmann worldsheet field: the string moves in one dimensional
space. 
The string bit degrees of freedom are specified by bosonic  
$a_\alpha^\beta$, ${\bar a}_\alpha^\beta=(a_\beta^\alpha)^\dagger$ and
fermionic $b_\alpha^\beta$, ${\bar b}_\alpha^\beta=(b_\beta^\alpha)^\dagger$
$N\times N$ matrix operators satisfying (anti)commutation relations
\bea
{}[a_\alpha^\beta,{\bar a}_\gamma^\eta]=\delta_\alpha^\eta\delta_\gamma^\beta,
\qquad \{b_\alpha^\beta,{\bar b}_\gamma^\eta\}=\delta_\alpha^\eta\delta_\gamma^\beta .
\eea
The bit number ${\cal M}=\tr({\bar a}a+{\bar b}b)$ is identified
with $P^+= m{\cal M}$,
and the dynamics is given by the Hamiltonian (to be related to $P^-$)
\bea
H&=&\frac{T_0}{2mN}\tr\left[({\bar a}^2 -i{\bar b}^2)a^2
-({\bar b}^2-i{\bar a}^2)b^2+({\bar a} {\bar b}
+{\bar b} {\bar a})ba+({\bar a} {\bar b}
-{\bar b} {\bar a})ab\right]
\label{susybitham}
\eea
$H$ has been chosen to commute with the supercharge $Q=\tr[{\bar a}b
e^{i\pi/4}+{\bar b}ae^{-i\pi/4}]$, which satisfies
$Q^2={\cal M}$, and so respects a supersymmetry.
We have written the coefficient as $T_0/m$ with $T_0$ the rest tension
of the emergent string for which $m$ will disappear as a parameter.
Later, when we study thermal perturbation theory, we will take 
$g=T_0/(2m\sqrt{2})$ as the expansion parameter. Here it is important
that $m$ and $g$ are independent parameters at the level of string bits.
We shall work directly with this
Hamiltonian in analyzing the high temperature behavior of the system,
which is best described in terms of the fundamental string bits.

The eigenstates of $H$ in the color singlet sector 
were obtained in \cite{sunthorn} in the limit $N\to\infty$. 
These $N=\infty$ eigenstates can be pictured as containing several
noninteracting (discretized) closed chains  of bits. A single closed chain
state is a linear combination of single trace states with fixed
bit number $M$. The ground single chain state has energy 
\bea
E_G&=&-\frac{1}{2}\sum_n\omega_n
=-\frac{T_0}{m}\cot\frac{\pi}{2M}=-\frac{2T_0M}{m\pi}+\frac{\pi T_0}{6Mm}
+{\mathcal O}(M^{-3})
\eea
where $\omega_n=(2T_0/m)\sin(n\pi/M)$, and in the last form 
we have taken $M$ large to show
the limit in which a chain becomes a continuous string. 
If we identify
\bea
P^-&\equiv&\frac{2T_0}{m\pi}{\cal M}+H,
\eea
the dispersion relation $P^-(P^+)$ is Lorentz invariant in 1+1 
dimensional spacetime,
in the limit $M\to\infty$: $2P^+P_G^-= \pi T_0/3$.

In this simplest string bit model the excited
states of a single closed chain
are those of left and right moving statistics 
waves \cite{bergmantclust,thornsubstructure} described in
the emergent string theory by a Grassmann
worldsheet field.
A wave in the $n$th normal mode adds energy $\omega_n$ to the ground state. 
If mode number $n<M/2$
is left moving then mode number $M-n$ is right moving. These
two modes have the same frequency $\omega_n$. 
The mode number $n$ takes on the
values $0,1,2,\ldots, M-1$. The zero mode $n=0$ is a fermionic operator,
whose square is unity,
which converts a state satisfying Bose-Einstein statistics to one satisfying
Fermi-Dirac statistics or vice versa. There is a cyclic constraint on the
occupied modes $\{n_i\}$, which is that $\sum_in_i$ is a multiple of $M$ if $M$
is odd, but it is an odd multiple of $M/2$ if $M$ is even. This mismatch
of cyclic constraints for $M$ even and odd is due to the fact that the
number of fermionic bits $b$ is odd (1 in this model). 
In the limit of continuous string
the cyclic constraint reduces to the familiar $L_0={\tilde L}_0$
constraint of closed string theory.

The energy of states with several closed strings is simply the sum of the
energies of the individual closed strings, reflecting the absence of
interactions between them when $N=\infty$. All of these multistring
states are color singlets and all have finite $P^-$ in the limit
$m\to0$ with $P^+=mM$ fixed. As noted
in \cite{sunthorn} the color nonsinglet states have $P^-$ of order $T_0/m$. 
Thus one has true color confinement in the limit $m\to0$ since in that limit
the only finite energy states are the color singlets with $M=\infty$.
If $m$ is kept finite but small, then the color non-singlets have
energy much greater than $\sqrt{T_0}$, and we can say we have effective
confinement.

At zero temperature the large $N$ limit is given by summing all
planar Feynman graphs. At finite temperature, the limit is given
by summing the planar graphs of thermal perturbation theory, reviewed
for the string bit system in the appendix. The canonical
partition function is given by\footnote{We have used $\tr$ to denote
the trace over matrix indices; here we use $\Tr$ to denote the thermal trace.} $Z=\Tr e^{-\beta P^0}$ where
\bea
P^0&=&\frac{P^++P^-}{\sqrt{2}}=\frac{1}{\sqrt{2}}
\left[\left(m+\frac{2T_0}{m\pi}\right){\cal M}+H\right]
\equiv \omega{\cal M}+\frac{H}{\sqrt{2}}\ .\eea
The vertices of the thermal graphs are determined by the
terms in $H$.
The sum of
connected graphs calculates $\ln Z$. This sum is easily shown to
have the structure
\bea
\ln Z=N^2f_0(\beta)+f_1(\beta)+\frac{1}{N^2}f_2(\beta)+\cdots
\label{oneovern}
\eea
where the leading term $f_0$ is found by calculating the sum of planar
diagrams.
The presence of the $N^2$ term is due to the fact that the 
operators $a, b$ are $N\times N$ matrices with $N^2$
elements. As we shall find, the calculation of $\ln Z$ using
the known energy spectrum at $N=\infty$, at low temperature
($\beta$ large) gives a contribution to $f_1(\beta)$ (because
color singlet states with large $M$ dominate). This contribution 
blows up when $\beta<\beta_H$,
predicting the ultimate temperature $\beta_H^{-1}$. In string models the
singularity in $f_1(\beta)$ is of the form $(\beta-\beta_H)^p$ where
the power $p=-\alpha-1$ 
depends on the details of the model. For the simple model
studied here we find $\alpha=-3/2$ implying $p=1/2$.
The corresponding
contribution to $f_0$ comes from color adjoint states and is suppressed
by factors of $e^{-\beta T_0/m}$. In the limit of absolute
confinement $m\to0$, this implies that $f_0(\beta)=0$ at
low temperature. On the other hand, calculating with the graphical
expansion shows no problem with arbitrarily high temperatures which
are dominated by the string bit description suggesting there
is no limiting temperature. For these two
facts to be compatible, the ultimate temperature must be an
artifact of the large $N$ limit. At finite $N$ the string bit system
has a finite number of degrees of freedom, and hence
$\ln Z$ should be a smooth function of $\beta$ for the whole range 
$0<\beta<\infty$.

In this note we shall calculate, from the $N=\infty$ eigenvalues of $H$,
the value of $\beta_H$ and also determine the power $p=1/2$. Then
$f_1(\beta)=-K_1(\beta-\beta_H)^{1/2}+K_2$ for $\beta$ slightly larger
than $\beta_H$. It then follows that 
$f_1^{\prime\prime}\sim(K_1/4)(\beta-\beta_H)^{-3/2}$. Since $\partial^2\ln Z/\partial\beta^2>0$, we must have $K_1>0$. It is
natural to guess that at finite $N$, this function is made smooth by
the substitution $(\beta-\beta_H)^{-3/2}\to ((\beta-\beta_H)^2+\eta(N))^{-3/4}$,
where $\eta(N)$ is some function of $N$ which vanishes as $N\to\infty$.
With this ansatz one can then integrate back to determine that 
$\ln Z$ for $\beta$ near $\beta_H$ would be 
proportional to the function\footnote{More generally one could 
construct a family  $g_k(\beta,N)$, each with a different $\gamma_k$. Then a linear combination
of this family would remove the discontinuities of the $N=\infty$
limit in the same way.}
\bea
g(\beta,N)&=&\frac{N^2}{\sqrt{\gamma}}(\beta_H-\beta)
\frac{\Gamma(1/4)\sqrt{\pi}}{\Gamma(3/4)}+\int_0^\beta dt(\beta-t)
\left[(t-\beta_H)^2+\frac{\gamma^2}{N^8}\right]^{-3/4}\\
&\sim&\begin{cases}
-4(\beta-\beta_H)^{1/2}+4\beta_H^{1/2}-2\beta\beta_H^{-1/2}&\beta>\beta_H\\
\phantom{.}&\\
\frac{N^2}{\sqrt{\gamma}}(\beta_H-\beta)
\frac{\Gamma(1/4)\sqrt{\pi}}{\Gamma(3/4)}-4(\beta_H-\beta)^{1/2}+4\beta_H^{1/2}-2\beta\beta_H^{-1/2}&
\beta<\beta_H\end{cases}
\label{dehagedorn}
\eea
where the last lines show the large $N$ behavior. The determination
that $\eta(N)=\gamma^2/N^8$ is made by requiring that the divergence as
$N\to\infty$ be precisely proportional to $N^2$ as dictated by the rules of
the $1/N$ expansion. We have fixed the integration constants
$A\beta+B$ so that the $N^2$ term is absent when $\beta>\beta_H$.
We have not yet learned enough about $1/N$ corrections to confirm the 
validity of this ansatz, but if it is valid, the physical interpretation of
the $N^2$ term, which is present only for $\beta<\beta_H$, is that
it signals the liberation of the fundamental string bit degrees of
freedom.

In the following sections we discuss our results in detail. Section
2 gives a brief review of the Hagedorn phenomenon for a single
free string as it is described in lightcone parameterization.
Section 3 then extends the discussion to the string discretized
as a chain of string bits. We obtain the Hagedorn temperature as a function 
of the discretization unit $m$. Section 4 concludes the paper.
An appendix which reviews thermal perturbation theory, needed
in the high temperature analysis of Section 3, in the context of
string bit models is included at the end.

\section{The free lightcone string at finite temperature}
\subsection{A general ideal gas}
Consider a system of bosons of various species $b$ and fermions of
various species $f$. Here $b$ and $f$ can include momentum as well as
internal state labels. In the absence of interactions the canonical
partition function is
\bea
Z&=&\prod_b \frac{1}{1-e^{-\beta \epsilon_b}}\prod_f
(1+e^{-\beta \epsilon_f})\\
\ln Z&=&\sum_f\ln(1+e^{-\beta \epsilon_f})-\sum_b\ln(1-e^{-\beta \epsilon_b})\\
&=&\sum_{n=1}^\infty\frac{1}{n}\left[\sum_b e^{-n\beta\epsilon_b}
+(-)^{n-1}\sum_f e^{-n\beta\epsilon_f}\right]
\label{zideal}
\eea
We see that the gas partition function can be expressed in terms of
the partition functions for a single particle immersed in heat
baths of temperatures $\beta^{-1}, (2\beta)^{-1},
\ldots, (n\beta)^{-1}),\ldots$. More specifically the $n$th term involves
either the single particle partition function 
\bea
z(n\beta)&=&\sum_b e^{-n\beta\epsilon_b}+\sum_f e^{-n\beta\epsilon_f}
=\sum_k e^{-n\beta\epsilon_k},\qquad {\rm for}\quad n\quad{\rm odd}
\eea
or the single superparticle partition function
\bea
z^S(n\beta)&=&\sum_b e^{-n\beta\epsilon_b}-\sum_f e^{-n\beta\epsilon_f},
\qquad {\rm for}\quad n\quad{\rm even}
\eea
When the particle spectrum is supersymmetric, as in the model
studied here, $z^S=0$.
\subsection{Hagedorn temperature for the lightcone string}
The Hagedorn temperature is by definition the lowest temperature
above which the partition function of the system diverges. Assuming the
divergence does not come from the sum over $n$ in (\ref{zideal}), 
we see that the
the Hagedorn temperature satisfies $z(\beta_H-\epsilon)=\infty$, because
then all the $z((2n+1)\beta_H-\epsilon), z^S(2n\beta_H-\epsilon)$ 
for $n>0$ are finite. 
Thus to determine the Hagedorn temperature, it suffices to examine the partition
function for a single particle in a heat bath!

In the lightcone description the energy of a closed string is expressed
as
\bea
P^0&=&\frac{1}{\sqrt{2}}(P^++P^-),\qquad P^-
=\frac{4\pi T_0(L_0+{\tilde L}_0+1/12)}{2P^+}
\eea
where $L_0$ (${\tilde L}_0$) is the transverse
string mode number operator for left (right) moving waves. 
They  depend in detail
on the string model of interest. Here we assume the simplest possible
transverse dynamics, namely a single fermion field on a closed string
worldsheet (hence the $1/12$ in $P^-$ above). For more elaborate 
string models one simply adds more worldsheet fields. 

For all models, the physical states of a closed string 
satisfy the constraint $(L_0-{\tilde L}_0)\ket{\psi}=0$.  We 
must therefore insert the projection operator
\bea
{\cal P}_{\rm Phys}
=\int_0^{2\pi}\frac{d\theta}{2\pi} e^{i\theta(L_0-{\tilde L}_0)}
\eea
in the thermal trace.
The canonical partition function for a single string in a heat bath
at temperature $\beta^{-1}$ is then
\bea
z(\beta)&=&\int_0^\infty dP^+ \int_0^{2\pi}\frac{d\theta}{2\pi}\Tr e^{-\beta P^0}
 e^{i\theta(L_0-{\tilde L}_0)}
=\int_0^\infty dP^+\int_0^{2\pi}\frac{d\theta}{2\pi}\Tr e^{-(\beta/\sqrt{2})(P^++P^-)}e^{i\theta(L_0-{\tilde L}_0)}\nonumber\\
&=&\int_0^\infty dP^+ e^{-(\beta/\sqrt{2})[P^++\pi T_0/(6P^+)]}
\int_0^{2\pi}\frac{d\theta}{2\pi}
\prod_{n=1}^\infty \left|1+e^{-\beta\pi T_0n\sqrt{2}/P^++in\theta}\right|^2
\eea
The Hagedorn temperature is the temperature above which the
integral over $P^+$ diverges. Putting $z=e^{-\beta\pi T_0\sqrt{2}/P^++i\theta}$,
we see that $z\to e^{i\theta}$ for $P^+\to\infty$ and the product
$\prod_n(1+z^n)(1+z^{*n})$ will be maximized in this limit for 
$\theta=0$.
To find $\beta_H$ in this
simple model we need the $z\to1$ behavior of the product
$\prod_n|1+z^n|^2$. One can either change variables via the Jacobi
imaginary transformation, or for the
leading behavior as $z\to1$, it is enough to write
\bea
\ln \prod_n(1+z^n)&=&\sum_{n=1}^\infty\ln(1+z^n)
=\sum_{k=1}^\infty\frac{(-)^{k-1}}{k}\frac{z^{k}}{1-z^k}\nonumber\\
&\sim&\frac{1}{1-z}\sum_{k=1}^\infty\frac{(-)^{k-1}}{k^2}
=\frac{\pi^2}{12}\frac{1}{1-z}\to\frac{\pi^2}{12}\frac{P^+}{\beta\pi
T_0\sqrt{2}}
\eea
where the last form inserted the value of $z$ at $\theta=0$
and large $P^+$ for our model.
It is evident that the large $P^+$ behavior of the $P^+$
integrand is of the form $(P^+)^\alpha e^{-h(\beta)P^+}$ where
the power $\alpha$ is determined to be $\alpha=-3/2$
by integrating $\theta$ in
the neighborhood of zero. Thus the Hagedorn
temperature is determined by
\bea
0&=&h(\beta_H)=\frac{\beta_H}{\sqrt{2}}-
\frac{\pi}{12}\frac{\sqrt{2}}{\beta_H T_0},\qquad
\beta_H=\sqrt{\frac{\pi}{6T_0}}
\eea
Because $\alpha=-3/2<-1$, $z(\beta_H)$ is actually finite for this model,
the Hagedorn singularity being a square root branch point ($p=1/2$).
\section{Superstring bit model at finite temperature}
\subsection{Low temperature behavior at $N=\infty$}
The color singlet eigenstates of the Hamiltonian
(\ref{susybitham}) at $N=\infty$ were obtained in \cite{sunthorn}.
They are states of multi-closed-chains
with $P^+=Mm$ each with fermion worldsheet fields for which the normal
mode frequencies are $\omega_n= (2T_0/m)\sin(n\pi/M)$. The cyclic constraint
can be imposed through the projection operator
\bea
{\cal P}&=&\frac{1}{M}\sum_{k=0}^{M-1}(-)^{k(M-1)}e^{2\pi i k{\cal N}/M}
\eea
where ${\cal N}$ is the mode number operator with values $\sum_i n_l$
on a state with modes $n_i$ occupied.
Then the partition function for a single chain in a heat bath at temperature
$\beta^{-1}$ is
\bea
z(\beta)&=&\sum_{M=1}^\infty e^{-\beta E_G}\frac{1}{M}\sum_{k=0}^{M-1}
(-)^{k(M-1)}\prod_{n=1}^{M-1} \left(1+e^{-(\sqrt{2}\beta T_0/m)\sin(n\pi/M)
+2i\pi nk/M}\right)\\
E_G&\equiv&\frac{mM+P^-_G}{\sqrt{2}}=\frac{1}{\sqrt{2}}\left(
mM+\frac{2MT_0}{m\pi}-\frac{T_0}{m}\cot\frac{\pi}{2M}\right)
\eea
Multi-chain states can then be included as usual by including the
$z((2n+1)\beta)$ and $z^S(2n\beta)$ terms of the ideal gas formula for
$\ln Z$.

The summand is maximized by the $k=0$ term, so
the Hagedorn temperature for this string bit model is given by
the condition
\bea
0&=&\frac{\beta_H m}{\sqrt{2}}-\lim_{M\to\infty}
\frac{1}{M}\sum_{n=1}^{M-1}
\ln\left(1+e^{-(\sqrt{2}\beta_H T_0/m)\sin(n\pi/M)}\right)
\eea
To analyze this condition, it is convenient to define the variable
$\xi=\sqrt{2}\beta_H T_0/m$ and then rewrite the equation
as a formula for the discretization unit $m$ as a function of $\xi$:
\bea
\frac{m^2}{2 T_0}&=&\lim_{M\to\infty}
\frac{1}{M\xi}\sum_{n=1}^M
\ln\left(1+e^{-\xi\sin(n\pi/M)}\right)\nonumber\\
&=&\frac{1}{\xi}\int_0^1 dx
\ln\left(1+e^{-\xi\sin(x\pi)}\right)=\frac{2}{\xi}\int_0^{1/2} dx
\ln\left(1+e^{-\xi\sin(x\pi)}\right)
\eea
Then one should choose $\xi$ so that $m^2/T_0\ll1$ to compare to the
continuous string. Evidently $m\to0$ when $\xi\to\infty$. So to
recover the continuous string result we need the large $\xi$ behavior
of the integral on the right side
\bea
2\int_0^{1/2} dx
\ln\left(1+e^{-\xi\sin(x\pi/2)}\right)&=&
\frac{2}{\pi}\int_0^1 \frac{du}{\sqrt{1-u^2}}\ln(1+e^{-\xi u})\nonumber\\
&=&\frac{2}{\pi}\frac{1}{\xi}\sum_{k=1}^\infty \frac{(-)^{k-1}}{k^2}+O(\xi^{-3})
=\frac{\pi}{6\xi}+O(\xi^{-3})
\eea
so that
\bea
\frac{m^2}{2 T_0}&\sim&\frac{\pi}{6\xi^2},\qquad \xi\to\infty\\
\beta_H&=&\frac{m\xi\sqrt{2}}{2T_0}\sim \frac{\xi\sqrt{2}}{2T_0}\sqrt{
\frac{2\pi T_0}{6\xi^2}}=\sqrt{\frac{\pi}{6T_0}}
\eea
in agreement with the direct continuum calculation.

To assess the consequences of discreteness one can simply take $\xi$
finite. If $\xi$ is large $m$ will be small and the effects of discreteness
will be small. For example put $\xi=10$ and $M=50000$ and calculate:
\bea
\frac{m}{\sqrt{T_0}}&=&0.1029839985,\qquad \beta_H\sqrt{T_0}
=0.7282068418=1.006364824\sqrt{\frac{\pi}{6}}
\eea
For $\xi=2$ and $M=50000$ the numbers become
\bea
\frac{m}{\sqrt{T_0}}&=&0.52988823849,\qquad \beta_H\sqrt{T_0}
=0.7493668536=1.035607454\sqrt{\frac{\pi}{6}}
\eea
It appears that the value of the Hagedorn temperature is rather
insensitive to the discreteness parameter $m$!

We should stress that these results depend on $N=\infty$. We have also
restricted the partition sum to color singlet states. Non-singlet states
would add  a positive amount and certainly couldn't remove the singularity.
But the energies of the nonsinglet states are of order $T_0/m$ and so
are highly suppressed (when $m^2\ll T_0$) at low temperatures.

In our study of thermal perturbation theory, the expansion parameter
$g=T_0/(2m\sqrt{2})$ will be small in the opposite limit $m^2\gg T_0$.
In that case we need to put $\xi\ll1$, for which
\bea
\frac{m^2}{2T_0}&\sim& \frac{\ln2}{\xi},\qquad \xi\ll1
\eea
Then we find the Hagedorn temperature $\beta_H\sim \sqrt{\xi\ln2}/\sqrt{T_0}$.
In this limit $\xi\sim(g^2/T_0)16\ln2$. Thus the Hagedorn temperature
goes to $\infty$ in the limit of $g=0$: 
\bea
\beta_H&\sim& \frac{4g\ln2}{T_0}.= \frac{\sqrt{2}\ln2}{m}
\eea
\subsection{High temperature}
In the high temperature limit, we expect the fundamental constituents to
play an active visible role. We shall use thermal perturbation theory, reviewed
in the appendix, to analyze this limit. Write the energy of the
string bit system as
\bea
P^0&=&\frac{1}{\sqrt{2}}\left[\left(m+\frac{2T_0}{m\pi}\right){\cal M}+H\right]
\label{pzero}
\eea
so that in the notation of the appendix $\omega=(m+2T_0/(m\pi))/\sqrt{2}$
and $g=T_0/(2m\sqrt{2})$ is the expansion parameter. The small expansion
parameter condition $g\ll\omega$
translates to $m^2\gg T_0$ which is the opposite of the continuous string
limit we are eventually interested in, but we hope some of our qualitative
insights will retain some validity, at least at high temperature.

Even if $g$ is assumed small, the high temperature limit requires at
least partial summation to all orders, because the bare
boson propagators blow up as $(\beta\omega)^{-1}$ as $\beta\to0$ making
successive terms in perturbation theory blow up more and more
severely. In the appendix we show that the solution to a one loop
Dyson equation for the boson propagator reduces the divergence to
$(2\beta g)^{-1/2}$ which is sufficient to make the remaining terms in the
expansion finite as $\beta\to0$. The upshot is that
the zeroth order high temperature behavior of the
partition function is modified from $(\beta\omega/2)^{-N^2}$
to  $(\beta g)^{-N^2/2}$. Some support for believing this result
captures the qualitative small $\beta$ behavior is that the
power of $\beta$ (though not the constant) agrees with the known
exact solution of the $N=1$ case.
\section{Concluding Remarks}
In this note we have analyzed the Hagedorn phenomenon in string bit
models. We started with a derivation of the Hagedorn temperature
of the continuous string from the point of view
of lightcone quantization, which is perhaps less familiar than
other treatments. Then we basically repeated this derivation for
the simplest string bit model. In this case the Hagedorn temperature
depends on $m$ the discrete unit of $P^+$. It ranges from $\sqrt{\pi T_0/6}$
for $m/\sqrt{T_0}\to0$ to $\infty$ for $m/\sqrt{T_0}\to\infty$. We also
analyzed the high temperature behavior of the system employing
thermal perturbation theory. In string bit models the Hagedorn
phenomenon cannot reflect a phase transition at finite $N$. We 
presented a possible hypothesis (\ref{dehagedorn}) for the
behavior of $\ln Z$ near the Hagedorn temperature, which illustrates
how a perfectly smooth function of temperature at finite $N$
induces the Hagedorn phenomenon when $N\to\infty$. Namely, in that
limit the leading term $N^2f_0$ is present only at temperatures
above the Hagedorn temperature. The string bit degrees of freedom
start to become more active, but in a smooth way, above and near the 
Hagedorn temperature. So far (\ref{dehagedorn}) is merely an educated guess--
it needs to be tested by studying higher orders in the $1/N$
expansion.
\vskip14pt
\noindent\underline{Acknowledgments}:
This research was supported in part by the Department
of Energy under Grant No. DE-FG02-97ER-41029.
\appendix
\section{Perturbation Theory}
\begin{figure}[ht]
\begin{center}
\hskip1in\includegraphics[height=1in]{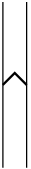}\hskip2.5in
\includegraphics[width=1in]{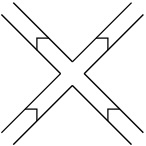}
\newline
\newline
 (a)\hskip3in(b)
\caption{Graphical rules for thermal perturbation theory. The
arrow points from an ${\bar a}$ to an $a$. Operators ordered
left to right are represented by graph elements ordered top to bottom.
Thus the propagator (a) contributes a factor $(e^{\beta\omega}-1)^{-1}$ if its 
arrow points down and contributes the factor
$e^{\beta\omega}(e^{\beta\omega}-1)^{-1}$ if its arrow points up. 
In a graph contributing to $n$th order
the top to bottom order of the vertices (b) coincides with the left to
right ordering of the $n$ perturbation operators.}
\label{propvertex}
\end{center}
\end{figure}

\begin{figure}[ht]
\begin{center}
\hskip.75in\includegraphics[width=1in]{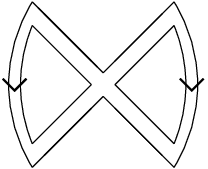}\hskip2in
\includegraphics[width=1.5in]{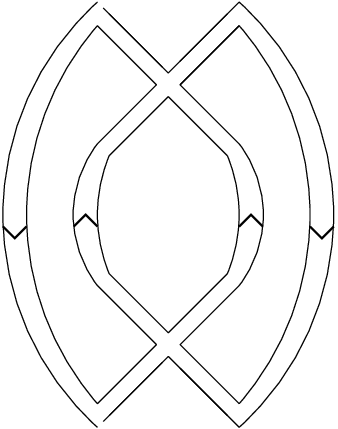}
\newline
\newline
 (a)\hskip3in(b)
\caption{Example of graphs contributing to the perturbation
expansion of the partition function. Its logarithm $\ln Z$ is calculated
by restricting to connected graphs. These particular graphs, being
planar contribute to leading order in the large $N$ limit.}
\label{12graphs}
\end{center}
\end{figure}

Here we give a brief review of thermal perturbation theory taking
advantage of the special features of string bit models. We concentrate on
the simplest model defined by (\ref{pzero}).
\bea
P^0=\omega{\cal M}+\frac{g}{N}\tr{\cal V} 
\eea
where ${\cal V}={\bar a}{\bar a}aa+\cdots$ are the quartic operators
within the square brackets of (\ref{susybitham}). We develop the
perturbation expansion as a power series in $g$.
\bea
Z&=&\Tr e^{-\beta P^0}=\sum_{n=0}^\infty
\frac{1}{n!}\Tr e^{-\beta\omega{\cal M}}\left(\frac{-\beta g}{N}
\tr{\cal V}\right)^n\\
&=&\left(\frac{1+e^{-\beta\omega}}{1-e^{-\beta\omega}}\right)^{N^2}
\sum_{n=0}^\infty
\frac{1}{n!}\VEV{\left(\frac{-\beta g}{N}\tr{\cal V}\right)^n}\\
\VEV{\Omega}&\equiv&\left(\frac{1+e^{-\beta\omega}}{1-e^{-\beta\omega}}
\right)^{-N^2}\Tr e^{-\beta\omega{\cal M}}\Omega
\eea
where $\tr$ denotes the matrix trace and $\Tr$ denotes the thermal trace.
The average is computed by applying Wick's theorem with the contraction rules
\bea
\VEV{{\bar a}_\alpha^\beta a_\gamma^\kappa}&=&\delta_\alpha^\kappa
\delta_\gamma^\beta\frac{1}{e^{\beta\omega}-1},\qquad
\VEV{a_\gamma^\kappa{\bar a}_\alpha^\beta }=\delta_\alpha^\kappa
\delta_\gamma^\beta\frac{e^{\beta\omega}}{e^{\beta\omega}-1}\\
\VEV{{\bar b}_\alpha^\beta b_\gamma^\kappa}&=&\delta_\alpha^\kappa
\delta_\gamma^\beta\frac{1}{e^{\beta\omega}+1},\qquad
\VEV{b_\gamma^\kappa{\bar b}_\alpha^\beta }=\delta_\alpha^\kappa
\delta_\gamma^\beta\frac{e^{\beta\omega}}{e^{\beta\omega}+1}
\eea
A graphical representation of Wick's theorem is constructed using
't Hooft's double line notation for the matrix operators. The
propagator and vertex are shown in Fig.~\ref{propvertex}.
First and second order examples of the application of the graphical rules
are shown in Fig.~\ref{12graphs}.

To define the $1/N$ expansion, we calculate the sum of
connected graphs which gives the perturbation expansion for
$\ln Z$. Then restricting the sum to planar diagrams give the leading order as
$N\to\infty$, which is of order $N^2$. Both diagrams 
shown in Fig.~\ref{12graphs} are planar and contribute to leading order.
The structure of the $1/N$ expansion in string bit models is that
shown in (\ref{oneovern}).

Since the boson propagators blow up at high temperature ($\beta\to0$), 
this limit is not amenable to straight perturbation theory. At least
some form of (partial) summation to all orders must be attempted.
A relatively simple partial summation is to set up a Dyson equation
for the ``self-energy'' $\Pi$ defined as the one particle
irreducible (1PIR)  two point function, in terms of which the
fully corrected propagator $\Delta$ is, at high temperature,
\bea
\Delta&=&\frac{1}{{\beta\omega}}\sum_{n=0}^\infty
\frac{\Pi^n}{(\beta\omega)^n}=\frac{1}{\beta\omega-\Pi}
\eea
At finite temperature the formalism is complicated by the distinction
that must be kept between $\VEV{{\bar a}a}\neq\VEV{a{\bar a}}$.
At high temperature, this distinction disappears
and both propagators are approximately $(\beta\omega)^{-1}$.
Here we limit the discussion to this simplifying limit. In general
both $\Delta$ and $\Pi$ are matrices in internal color space.
There are two cases where they can be treated as numbers: namely
$N=1$ when they {\it are} numbers, and $N=\infty$ when the indices are simply
spectators, which factor out of both sides of the Dyson equation.

In the latter case ($N=\infty$)
the one loop Dyson equation for the boson propagator 
reads (for $N=1$ change the 2 to a 4)
\bea
\Pi&=&-2\beta g\frac{1}{\beta\omega-\Pi}
\eea
which is a quadratic equation for $\Pi$:
\bea
0&=&\Pi^2-\Pi(\beta\omega)-2\beta g\\
\Pi&=&\frac{1}{2}\left[\beta\omega-\sqrt{\beta^2\omega^2+8\beta g}\right]
\eea
where the branch of the square root is chosen so that $\Pi\to0$
at $g=0$. We see that in the high temperature limit $\Pi\sim-\sqrt{2\beta g}$.
In this approximation, the high temperature behavior of $\ln Z $ is
\bea
\ln Z&\sim&-N^2\ln(1-e^{-\beta\omega})
+N^2\sum_{n=1}^\infty\frac{1}{n}\left(\frac{\Pi}{\beta\omega}\right)^n
\nonumber\\
&\sim&-N^2\ln(\beta\omega)-N^2\ln\left(1-\frac{\Pi}{\beta\omega}\right)
=-N^2\ln\frac{\beta\omega+\sqrt{\beta^2\omega^2+8\beta g}}{2}\nonumber\\
&\sim& -\frac{N^2}{2}\ln(2\beta g)
\eea
The effect of the interactions in this approximation is to soften
the singularity in $Z$ at $\beta\to0$ but not to remove it entirely.

Of course this only takes into account a subset of the terms
in the perturbation expansion, after which the terms in perturbation expansion
are at least finite as $\beta\to0$.
But it would be helpful to test how
it does in a context where the exact answer is known. The special case
$N=1$ is an instructive example. For simplicity we drop the
fermionic operators. In that case the Hamiltonian is a
function of the number operator ${\cal M}={\bar a}a$:
\bea
H={\cal M}\omega+g{\bar a}{\bar a}aa={\cal M}\omega+g({\cal M}^2-{\cal M})
\eea 
and hence
\bea
Z&=&\sum_{n=0}^\infty e^{-n\beta\omega-\beta g(n^2-n)}
\eea
Taking the limit $\beta\to0$ the sum can be approximated by an integral over
a variable $x=n\sqrt{\beta}$:
\bea
Z&\sim&\frac{1}{\sqrt{\beta}}\int_0^\infty dx e^{-x\sqrt{\beta}\omega
-g(x^2-x\sqrt{\beta})}\approx\frac{1}{\sqrt{\beta}}\int_0^\infty dx
e^{-gx^2}=\frac{\sqrt{\pi}}{2\sqrt{\beta g}}
\eea
We see that the one loop approximation gets the power of $\beta$
right but not the constant prefactor.

It is also encouraging that this last result can be obtained by
analyzing the perturbation expansion for $Z$. At order $n$, one can count
the number of Wick contraction schemes of $\VEV{({\bar a}^2a^2)^n}$
and find that there are exactly $(2n)!$. Since at $N=1$ and
high temperature all graphs at each order are equal, we can conclude that
\bea
Z&=&\frac{1}{1-e^{-\beta\omega}}\sum_{n=0}^\infty\left(\frac{-\beta g}{\beta^2\omega^2}\right)^n\frac{(2n)!}{n!}
\eea
a sum that has zero radius of convergence. However we can use the Borel 
summation trick
\bea
(2n)!=\int_0^\infty dt t^{2n}e^{-t}
\eea
to interpret the sum as
\bea
Z&=&\frac{1}{1-e^{-\beta\omega}}\int_0^\infty dt e^{-t-gt^2/(\beta\omega^2)}
=\frac{1}{1-e^{-\beta\omega}}\sqrt{\frac{\beta\omega^2}{g}}
\int_0^\infty dt e^{-t\omega\sqrt{\beta/g}-t^2}\\
&\sim& \frac{1}{2}\sqrt{\frac{\pi}{\beta g}},\qquad {\rm as}\quad \beta\to0
\eea
Thus a complete analysis of perturbation theory in this simple case gets
the power of $\beta$ {\it and} the prefactor right. If we could
get an accurate count of the number of planar connected vacuum diagrams
at each order $n$, we could make a similar statement about high 
temperature in the large $N$ limit. Our one loop Dyson equation evaluation
provides support for the high temperature behavior 
$\ln Z\to(N^2/2)\ln(\beta g)+C$ in
the $N\to\infty$ limit but gives no reliable 
nonperturbative information about the constant $C$.

Because fermion propagators are perfectly finite as $\beta\to0$,
the presence of fermion lines in the graphical rules does not affect the
singular high temperature behavior of the partition function,
which is entirely determined by the bosonic degrees of freedom. However,
they will certainly contribute to the subleading behavior and in particular can
be expected to contribute to the constant $C$.


\end{document}